\documentstyle[aps,prl,twocolumn,epsf]{revtex}
\preprint{NUC-MINN-01/...-T}

\newcommand{\be}{\begin{equation}} 
\newcommand{\ee}{\end{equation}}
\newcommand{\ba}{\begin{eqnarray}}
\newcommand{\ea}{\end{eqnarray}}

\title{Spontaneous Symmetry Breaking with Abnormal Number of
Nambu-Goldstone Bosons and Kaon Condensate}

\author{V.A.~Miransky$^{*}$}

\address{Department of Applied Mathematics, University of Western
Ontario, London, Ontario N6A 5B7, Canada}

\author{I.A.~Shovkovy$^{*}$}

\address{School of Physics and Astronomy, University of Minnesota, 
Minneapolis, MN 55455, USA}

\draft
\begin{document}

\maketitle

\begin{abstract}
We describe a class of relativistic models incorporating a finite density
of matter in which spontaneous breakdown of continuous symmetries leads to
a lesser number of Nambu-Goldstone bosons than that required by the
Goldstone theorem. This class, in particular, describes the dynamics of
the kaon condensate in the color-flavor locked phase of high density QCD.
We describe the spectrum of low energy excitations in this dynamics and
show that, despite the presence of a condensate and gapless excitations,
this system is not a superfluid.  
\end{abstract}




The Goldstone theorem is a cornerstone of the phenomenon of spontaneous
breakdown of continuous global symmetries. It is applicable both to
relativistic field theories with exact Lorentz symmetry \cite{GSW} and to
most condensed matter systems \cite{BL} where there is no this symmetry.
However, there is an important difference between these two cases. While
in Lorentz invariant systems, the Goldstone theorem is universally valid,
it is not so in condensed matter systems. For example, it does not apply
to condensed matter systems with long range interactions \cite{BL}. From
the technical viewpoint, the difference is connected with a kinetic term,
and derivative terms, in general, in a Lagrangian density: while their
form is severely restricted by the Lorentz symmetry, it is much more
flexible in systems where this symmetry is absent.

In this letter, we describe the phenomenon of spontaneous symmetry
breaking of continuous symmetries with an abnormal number of
Nambu-Goldstone (NG) bosons taking place at a sufficiently high density of
matter in a class of models {\it without} long range interactions. Here by
``abnormal", we understand that the number of gapless NG bosons is less
than the number of the generators in the coset space ${G/H}$, where $G$ is
a symmetry of the action and H is a symmetry of the ground state. On the
other hand, as we will see, the degeneracy of the ground state retains
conventional: it is described by transformations connected with all the
generators from the coset space. It is noticeable that this class of
models describes a recently suggested \cite{S,BS,KR} scenario with a kaon
condensate in the color-flavor locked (CFL) phase of high density QCD
\cite{ARW}.

We will illustrate this phenomenon in a toy model with the following
Lagrangian density:
\ba
{\cal L} &=& (\partial_{0} +i \mu )\Phi^{\dagger}
(\partial_{0} -i \mu )\Phi \nonumber \\
&-&v^2\partial_{i}\Phi^{\dagger}
\partial_{i}\Phi -m^{2}\Phi^{\dagger} \Phi
-\lambda(\Phi^{\dagger} \Phi)^{2},
\label{L-model}
\ea
where $\Phi$ is a complex doublet field and $v$ is a velocity parameter.
Since here the Lorentz symmetry is broken by the terms with the chemical
potential $\mu$, the velocity $v \leq 1$ in general. The chemical
potential $\mu$ is provided by external conditions (to be specific, we
take $\mu > 0$) \cite{footn}.  The above Lagrangian density is invariant
under global $SU(2)\times U(1)$. The $SU(2)$ will be treated as the
isospin group $I$ and the $U(1)$ will be associated with hypercharge $Y$.
The electric charge is $Q = I_{3} + Y$. This model describes the essence
of the dynamics of the kaon condensate \cite{BS} (see below).

When $\mu<m$, it is straightforward to derive the tree level spectrum of
the physical degrees of freedom. To this end, we switch to the momentum
space by decomposing all four real components of $\Phi$ field in plane
waves. Then, the quadratic part of the above Lagrangian density takes the
following form:
\ba
{\cal L}^{(2)}(\omega,q) &=& 
\frac{1}{2}
\left(\begin{array}{cc}
\phi_{1}^{*} & \phi_{2}^{*}
\end{array}\right)
{\cal M}
\left(\begin{array}{c}
\phi_{1} \\ \phi_{2}
\end{array}\right) \nonumber \\
&+& \frac{1}{2} \left(\begin{array}{cc}
\tilde{\phi}_{1}^{*} & \tilde{\phi}_{2}^{*}
\end{array}\right)
\tilde{\cal M}
\left(\begin{array}{c}
\tilde{\phi}_{1} \\ \tilde{\phi}_{2}
\end{array}\right),
\label{qdr}
\ea
where the real and imaginary parts of each component of the doublet 
were introduced, $\Phi^{T}=\frac{1}{\sqrt{2}} (\phi_{1}+i\phi_{2},
\tilde{\phi}_{1} +i\tilde{\phi}_{2})$. Note that their Fourier transforms
satisfy $\phi_{i}^{*}(\omega,\vec{k})=\phi_{i}(-\omega,-\vec{k})$ and
$\tilde{\phi}_{i}^{*}(\omega,\vec{k})=
\tilde{\phi}_{i}(-\omega,-\vec{k})$. The matrices ${\cal M}$ and
$\tilde{\cal M}$ in Eq.~(\ref{qdr}) read
\be
\left(\begin{array}{cc}
\omega^{2}+\mu^{2}-m^{2}-v^{2}q^{2} & 2i\mu\omega \\
-2i\mu\omega & \omega^{2}+\mu^{2}-m^{2}-v^{2}q^{2} 
\end{array}\right).
\ee
The dispersion relations of the particles are determined 
from the equation $\mbox{Det}({\cal M})=0$. Explicitly, this
equation reads:
\ba
\left[(\omega-\mu)^{2}-m^{2}-v^{2}q^{2}\right]
\left[(\omega+\mu)^{2}-m^{2}-v^{2}q^{2}\right]=0,
\ea
i.e., the particle's dispersion relations are:
\ba
\omega_{1} &=& \tilde{\omega}_{1} = 
\pm (\sqrt{m^{2}+v^{2}q^{2}} +\mu) ,\\
\omega_{2} &=& \tilde{\omega}_{2} =
\pm (\sqrt{m^{2}+v^{2}q^{2}} - \mu) .
\ea
Of course, the positive and negative values of energy correspond to
creation and annihilation of excitations, respectively. Henceforth we will
consider only positive eigenvalues: for our purposes, no additional
information contains in the eigenstates with negative eigenvalues.

Eqs. (5) and (6) imply that
the particle spectrum contains two doublets with the energy gaps
$m+\mu$ and $m-\mu$, respectively.  In dense quark matter, the first
doublet can be identified with $(K^{-},\bar{K}^{0})$ and the second one
with $(K^{+}, K^{0})$. It is important to note that the chemical potential
causes splitting of the masses of particles and their antiparticles. This
point will be crucial for reducing the number of NG bosons in the
asymmetric phase considered below. The splitting is intimately connected
with the fact that $C, CP$, and $CPT$ symmetries are explicitly broken in
this system. In second quantized theory, the complex field $\Phi$
describes creation and annihilation of $(K^{-}, \bar{K}^{0})$ and
$(K^{+},K^{0})$, respectively. This is quite unusual because the
corresponding dispersion relations of these two doublets are not
identical.

By studying the potential of the above model in tree approximation, we
could check that the perturbative ground state becomes a local maximum
when $\mu>m$ [see Eq.~(\ref{qdr}) with $\omega=q=0$]. At this point, the
system experiences an instability with respect to forming a condensate.
In the new phase, a vacuum expectation value, $\varphi_{0}$, of the field
$\Phi$ occurs,
\be
\Phi=\left(\begin{array}{c} 0 \\ \varphi_{0}
\end{array}\right)
+\frac{1}{\sqrt{2}}\left(\begin{array}{c}
\phi_{1}+i\phi_{2} \\
\tilde{\phi}_{1}+i\tilde{\phi}_{2}
\end{array}\right).
\ee
This choice of the ``neutral" direction of the vacuum expectation value
corresponds to the conventional definition of the electric charge, $Q =
I_{3} + Y$.

By requiring that the new ground state is a minimum of the potential, we
derive 
\be
\varphi_{0}^{2} = \frac{\mu^{2}-m^{2}}{2\lambda}
\ee
for the vacuum expectation value of the field. In this ground state, 
the initial $SU(2)\times U(1)$ spontaneously breaks down to $U(1)_{Q}$.

\begin{center}
\begin{figure}
\epsfxsize=8.0cm
\epsffile[88 14 388 190]{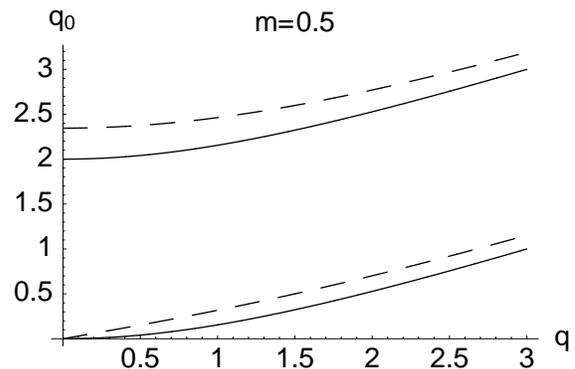}
\caption{Dispersion relations of four particles in the broken phase of
the model in Eq.~(\ref{L-model}). The solid and dashed lines represent 
the dispersion relations of charged and neutral states, respectively.
All quantities are given in units of $\mu$. To plot the figure we used
$m=0.5\mu$ and $v=1/\sqrt{3}$.}
\label{fig-disp-model}
\end{figure}
\end{center}

In the broken phase, the quadratic part of the Lagrangian density
looks formally the same as in Eq.~(\ref{qdr}). The matrices ${\cal M}$
and $\tilde{\cal M}$, however, are different. In particular, the first
one, describing two charged states, is
\be
{\cal M}=\left(\begin{array}{cc}
\omega^{2}-v^{2}q^{2}&2i\mu\omega \\
-2i\mu\omega&\omega^{2}-v^{2}q^{2} 
\end{array}\right),
\ee
while the other, describing two neutral states, is
\be
\tilde{\cal M}=\left(\begin{array}{cc}
\omega^{2}-2(\mu^{2}-m^{2})-v^{2}q^{2}&2i\mu\omega \\
-2i\mu\omega&\omega^{2}-v^{2}q^{2} 
\end{array}\right).
\ee
For the charged states, the dispersion relations are 
\ba
\omega_{1,2} =\sqrt{\mu^{2}+v^{2}q^{2}}\pm \mu.
\label{dsp1}
\ea 
For the neutral states, the dispersion
relations are given by
\be
\tilde{\omega}_{1,2} =\sqrt{3\mu^{2}-m^{2} +v^{2}q^{2}
\pm \sqrt{(3\mu^{2}-m^{2})^{2} + 4\mu^{2}v^{2}q^{2} }}.
\label{dsp2}
\ee
All four dispersion relations are shown in Fig.~\ref{fig-disp-model}.
As is easy to check, one of the charged states with the relation
\be
\omega_{2} =\sqrt{\mu^{2}+v^{2}q^{2}}- \mu,
\ee
and one of the neutral states with the relation 
\be
\tilde{\omega}_{2} =\sqrt{3\mu^{2}-m^{2} +v^{2}q^{2}
-\sqrt{(3\mu^{2}-m^{2})^{2} + 4\mu^{2}v^{2}q^{2} }},
\ee
describe NG bosons, i.e., gapless excitations whose energy goes to
zero as $q \to 0$ (see Fig.~\ref{fig-disp-model}). Indeed, in the far
infrared region, these relations take the following form:
\ba
\omega_{2} &\simeq & \frac{v^{2}q^{2}}{2\mu}, \label{NG} \\
\tilde{\omega}_{2} &\simeq & \sqrt{\frac{\mu^{2}-m^{2}}{3\mu^{2}-m^{2}}}v
q.
\label{NG1}
\ea
No other gapless states appear. Thus, there are two gapless NG bosons in
the system. It comes as a real surprise. Indeed, since the initial global
$SU(2)\times U(1)$ symmetry spontaneously breaks down to $U(1)_{Q}$, one
should expect the existence of three NG bosons. Where is the third one?!
To better understand the situation at hand, it is instructive to consider
the quadratic part of the potential around the ground state configuration
in the broken phase. For the two sets of the fields, the corresponding
quadratic forms are read from $-{\cal M}$ and $-\tilde{\cal M}$ where
$\omega=q=0$ is substituted. The eigenvalues of such matrices are
\ba 
\xi_{1,2} &=& \tilde{\xi}_{2} =0, \\
\tilde{\xi}_{1} &=& 2 (\mu^{2}-m^{2}).
\ea 
Therefore, we see that the potential part of the action has three flat
directions in the broken phase, as it should.  These three directions
would correspond to three NG bosons, related to the three broken
generators of the original $SU(2)\times U(1)$. The first order derivative
terms in the kinetic part of the action, however, prevent the appearance
of one charged, $K^{-}$, gapless mode. This fact is intimately connected
with the point that the presence of the chemical potential leads to 
splitting of the energy spectra of $K^{-}$ and $K^{+}$.

Another noticeable fact is that the energy $\omega_2$ of the $K^{+}$
gapless mode in Eq.~(\ref{NG}) is proportional to $q^2$ rather than to
$q$. This implies that, despite the presence of gapless modes, the Landau
criterion for superfluidity fails in this model \cite{foot-Landau}.
Therefore, this system is not a superfluid.

This toy model illustrates a rather general phenomenon. While in Lorentz
invariant systems with spontaneous breakdown of continuous symmetries the
degeneracy of the potential fixes the number of NG bosons being equal to
the number of generators $N_{G/H}$ in the coset space $G/H$, in systems
with a broken Lorentz symmetry the number of gapless NG bosons can be
lesser. The latter is connected with a form of the kinetic term which can
include first order derivative terms. In the systems under consideration,
the chemical potential (leading here to such first order derivative terms)
splits up energy gaps of charged particles and their antiparticles {\em
both} of which would be NG bosons otherwise.  On the other hand, a neutral
NG boson, a partner of a field with a nonzero vacuum expectation value,
always survives.  Therefore, the number of the gapless NG bosons reduces
by $N_{ch}$, where $N_{ch}$ is the number of charged particle-antiparticle
would be NG pairs. In the present model $N_{G/H}=3$ and $N_{ch}=1$, and as
a result there are only two (one neutral and one charged) NG bosons.

The choice of the simple model in Eq.~(\ref{L-model}) was not accidental
in this paper. It describes the essence of a much more complicated
dynamics of spontaneous symmetry breaking related to the kaon
condensation in the color-flavor locked (CFL) phase of dense quark matter
\cite{BS}. Let us turn to it.

We start by repeating briefly the analysis of Ref.~\cite{BS}. In the
chiral limit, the ground state of the three flavor dense quark matter
corresponds to the CFL phase \cite{ARW}. The original $SU(3)_{c}\times
SU(3)_{L}\times SU(3)_{R}$ symmetry of the microscopic action breaks down
to global ``locked" $SU(3)_{c+L+R}$ subgroup. The corresponding low energy
action for the NG bosons was derived in Refs.~\cite{CasGat,SonSt}. The
current quark masses break the original chiral symmetry of the model
explicitly. As a result, nonzero gaps appear in the spectra of the NG
bosons, and they become pseudo-NG bosons. Their dynamics could still be
described by the low energy action which, for sufficiently small current
quark masses, could be derived from the microscopic theory \cite{SonSt}.
By making use of an auxiliary ``gauge" symmetry, it was suggested in
Ref.~\cite{BS} that the low energy action of Refs.~\cite{CasGat,SonSt}
should be modified by adding a time-like covariant derivative to the
action of the composite field.

By neglecting a chemical potential of the electric charge, the low energy
effective Lagrangian density of Ref.~\cite{BS} (in Minkowski space) reads
\ba
{\cal L}_{eff} &=& \frac{f_\pi^2}{4} {\rm Tr}\left[
 \nabla_0\Sigma\nabla_0\Sigma^\dagger - v_\pi^2
 \partial_i\Sigma\partial_i\Sigma^\dagger \right]
\nonumber \\
&+&\frac{1}{2} \left[ (\partial_0 \eta^{\prime})^{2}
- v_{\eta^{\prime}}^2  (\partial_i\eta^{\prime})^{2} \right]
\nonumber \\
&+& 2 c\left[\det(M) \mbox{Tr} \left( M^{-1} \Sigma 
e^{\sqrt{\frac{2}{3}} \frac{i}{f_{\eta^{\prime}}} {\cal I} 
\eta^{\prime}}\right) + h.c.\right], 
\label{L-eff} \\
\nabla_0\Sigma &=& \partial_0 \Sigma
 + i \frac{M M^\dagger}{2p_F} \Sigma
 - i \Sigma \frac{ M^\dagger M}{2p_F} ,
\label{cov-der}
\ea
where $p_F$ is the quark Fermi momentum and $M$ is a quark mass matrix
chosen to be diagonal, i.e., $M=\mbox{diag}(m_{u},m_{d},m_{s})$. By
definition, ${\cal I}$ is a unit matrix in the flavor space, and $\Sigma$
is a unitary matrix field which describes the octet of the NG bosons,
transforming under the chiral $SU(3)_{L} \times SU(3)_{R}$ group as
follows:
\be 
\Sigma \to U_{L} \Sigma U_{R}^{\dagger}, 
\ee
where $(U_{L},U_{R}) \in SU(3)_{L} \times SU(3)_{R}$. In
Eq.~(\ref{L-eff}), we also took into account $\eta^{\prime}$ field which
couples to the octet when the quark masses are nonzero. The NG boson,
related to breaking the baryon number, was omitted, however.  Its
dynamics is not affected much by the quark masses.

One should notice from the definition of the covariant derivative in
Eq.~(\ref{cov-der}) that the combination of the quark mass matrices
$\mu_{eff}=MM^{\dagger}/2p_{F}$ produces effective chemical potentials
for different flavor charges. These chemical potentials are of dynamical
origin, and they are unavoidable.

The presense of the effective chemical potential $\mu_{eff}$ has far
reaching consequencies. In particular, if the mismatch of the quark
masses of different flavors is large enough [e.g., $m_{s}\agt
(\Delta^{2}m_{u})^{1/3}$], the perturbative CFL ground state becomes
unstable with respect to a kaon condensation.

The new ground state is determined by a ``rotated" vacuum expectation
value of the $\Sigma$ field, 
\ba
\Sigma_{\alpha}(\pi) &\equiv& \exp\left(i\alpha\lambda^{6}\right)
\exp\left(i\pi_{A}\lambda^{A}\right)  
\simeq  \exp\left(i\alpha\lambda^{6}\right) \nonumber \\ 
&\times& \left(
1 + \frac{i\pi_{A}\lambda^{A}}{f_\pi}
- \frac{\pi_{A}\pi_{B}\lambda^{A}\lambda^{B}}{2f_\pi^2} 
+ \ldots \right),
\label{new-gr-state} 
\ea 
where $\alpha$ is determined by requiring that the corresponding ground
state configuration is a global minimum of the potential energy. In the
simplest case with $m_{u}=m_{d}$, we derive
\be
\cos\alpha=\frac{4cp_{F}^{2}m_{u}(m_{s}+m_{u})}
{f_{\pi}^{2}(m_{s}^{2}-m_{u}^{2})^{2}}<1,
\label{alpha}
\ee
where (see Ref.~\cite{SonSt}) 
\be
c = \frac{3 \Delta^{2}}{2\pi^{2}}
\quad \mbox{and} \quad
f_{\pi}^{2} = \frac{21-8\ln2}{36} \frac{p_{F}^{2}}{\pi^{2}}.
\ee
The ground state with the kaon condensation which is determined by
Eq.~(\ref{new-gr-state}) breaks the $SU(2)\times U(1)_{Y}$ symmetry of
the effective action (\ref{L-eff}) down to $U(1)_{Q}$. This is exactly the
symmetry breaking pattern that we encountered in the toy model.

The derivation of the dispersion relations in the broken phase with the
kaon condensation involves rather tedious calculations. Qualitatively,
though, such a derivation is similar to that in the model in
Eq.~(\ref{L-model}). Additional difficulties come from more complicated
particle mixing. By omitting the details, we present the results.

With the choice of the vacuum expectation value in
Eq.~(\ref{new-gr-state}), the original 9 degrees of freedom ($\pi_{A}$
with $A=1,\ldots 8$ and $\eta^{\prime}$) group into two decoupled sets:
($\pi_{1}$, $\pi_{2}$, $\pi_{4}$, $\pi_{5}$) and ($\pi_{3}$, $\pi_{6}$,
$\pi_{7}$, $\pi_{8}\equiv\eta$, $\eta^{\prime}$).  The first set of states
contains all charged degrees of freedom (i.e., $\pi^{\pm}$ and $K^{\pm}$),
while the second set contains neutral ones (i.e., $\pi^{0}$, $K^{0}$,
$\bar{K}^{0}$, $\eta$ and $\eta^{\prime}$).

The dispersion relations for the charged degrees of freedom are 
straightforward to derive. Our
analysis shows that there is only 1 gapless NG boson in this set. In the
far infrared region $q\to 0$, its explicit dispersion relation reads 
\be
\omega \simeq \frac{q^{2}} {3\mu \cos\alpha(1+\cos\alpha)}, 
\ee 
where $\mu = (m_{s}^{2}-m_{u}^{2})/2p_{F}$. The analysis of the
dispersion relations of the other set is more difficult. But it is also
straightforward to show that there is only 1 gapless NG boson there as
well. Moreover, for $q\to 0$, the explicit dispersion relation reads
\be 
\omega \simeq
\frac{\sin\alpha}{\sqrt{2-7\cos^{2}\alpha+9\cos^{4}\alpha}}
\frac{q}{\sqrt{3}}. 
\ee 
We see that, as in the case of the simple model in Eq.~(\ref{L-model}),
there are only two gapless NG bosons, despite the fact that there are
three broken symmetry generators in the phase with the kaon condensate.
One should note, however, that as in the model in Eq.~(\ref{L-model}),
the effective potential obtained from (\ref{L-eff}) has the required
three flat directions. It is the first order derivative terms that are
responsible for producing a nonzero energy gap in the spectrum of one of
the charged bosons.  The corresponding opposite charge partner remains a
gapless NG boson. It is crucial to note, though, that its dispersion
relation behaves as $\omega \sim q^2$ for $q\to 0$. This implies that the
criterion of superfluidity is not satisfied in the phase of the dense
quark matter with kaon condensation \cite{foot-Landau}.

As has been argued in Refs.~\cite{BS,KR}, there is a good chance that the
phase with a kaon condensate may exist in a core of compact stars. Since
the spectrum of low energy excitations in this phase derived in the
present paper is very specific, implying in particular that the matter is
not superfluid, it could play an important role in detecting this phase.

In conclusion, in this paper the phenomenon of spontaneous symmetry
breaking with an abnormal number of NG bosons was described. It admits a
simple and clear interpretation. We expect that there exist wide
applications of this phenomenon that deserve further study. In passing,
we note that related systems in condense matter physics are ferromagnets
\cite{fer} and the superfluid $^{3}$He in the so-called A-phase
\cite{Volovik}.

\begin{acknowledgments}
We would like to thank G.E.~Volovik for bringing Ref.~\cite{Volovik} to
our attention.  The work of I.A.S. was supported by the U.S. Department of
Energy Grant No.~DE-FG02-87ER40328.
\end{acknowledgments}

\end{document}